\documentclass[12pt,british]{article}
\usepackage[T1]{fontenc}
\usepackage[latin9]{inputenc}
\usepackage{geometry}
\usepackage{babel}
\usepackage{verbatim}
\usepackage{mathrsfs}
\usepackage{amsmath}
\usepackage{amssymb}
\usepackage{stackrel}
\usepackage{graphicx}
\usepackage{setspace}
\usepackage[unicode=true,pdfusetitle,
 bookmarks=true,bookmarksnumbered=false,bookmarksopen=false,
 breaklinks=false,pdfborder={0 0 1},backref=false,colorlinks=false]
 {hyperref}

\makeatletter
\@ifundefined{date}{}{\date{}}
\usepackage{soul}

\makeatother

\begin{document}
\title{Critical phenomena of RNA-like polymers - a synopsis}
\author{R. Dengler \thanks{ORCID: 0000-0001-6706-8550}}
\maketitle
\begin{abstract}
This work examines field theories for RNA-like polymers with single
strand and double strand polymers and a periodic base sequence. These
field theories originate from lattice models, describe polymers in
a good solvent, and in principle exactly describe the critical behavior.
A central role is played by the conventional one-component branched
polymer and the mapping of the corresponding field theory to the Lee-Yang
field theory in two less dimensions. 

Critical phenomena in the context of polymers as well as the Lee-Yang
model entail pecularities, which we derive in detail. A new result
is that the critical point of RNA-like branched polymers (with periodic
base sequence) looks like the critical point of classical one-component
branched polymers, but with one more critical exponent for the single
strand polymer. A random base sequence generates additional relevant
interactions, and invalidates the simple picture.
\end{abstract}
\tableofcontents{}\pagebreak{}

\section{Introduction}

There is a natural connection between linear as well as branched polymers
in a good solvent and critical phenomena. One of the parameters of
such a system is polymer length $\ell$, and a large polymer length
usually entails a long correlation length $\xi\sim\ell^{\nu}$, where
$\nu$ is the usual critical exponent. In other words, $1/\ell$
plays the role of a temperature variable. The temperature variable
of lattice models and field theories for polymers is unphysical in
this sense, it acts as chemical potential for monomers and is a function
of polymer length. Another obvious connection are the Feynman diagrams
of polymer field theories, which topologically are identical with
polymer conformations. 

The formal mapping of the statistics of a linear polymer in a good
solvent to a field theory and to critical phenomena dates back to
1972, when de Gennes proved the equivalence to the $n=0$-limit of
the $O\left(n\right)$-symmetric $\varphi^{4}$ model \cite{Gennes1972}.
The effect of the $n\rightarrow0$-limit is to eliminate all closed
loops on a lattice, or on the level of Feynman diagrams. There remain
polymer chains extending from an external source to an external sink
(a magnetic field in $\varphi^{4}$ spin language). 

The simplest physical application corresponds to the limit of a weak
magnetic field, where there is a low concentration of non-interacting
polymers. The next level of complexity are solutions with a finite
concentration of linear polymers, which can be described within the
$\varphi^{4}$ field theory with a finite magnetic field \cite{Cloiseaux1975}.
Here, however, we always restrict ourselves to the single polymer
case, i.e. weak external fields.

\subsection{Branched polymers}

The next level of complexity is a branched polymer in a good solvent,
examined in the field theory context by Lubensky and Isaacson in 1979
\cite{Lubensky1979}. The model describes the statistics of lattice
animals (clusters on a lattice) and the related problem of the statistics
of branched polymers in the dilute limit. The upper critical dimension
is $8$.

The field theory used by these authors involves an $n\rightarrow0$
limit in a somewhat complicated way. A simplified version was written
down by Parisi and Sourlas \cite{Parisi1981}, who demonstrated the
equivalence to the Lee-Yang problem in two less dimensions with the
help of a hidden supersymmetry. The critical exponents of the Lee-Yang
system in one dimension are known exactly, and there follow predictions
for branched polymers in three dimensions. The predictions agree numerically
with the result of series expansions (``high temperature'' expansions)
and Monte Carlo simulations \cite{Parisi1981}, and the mapping appears
to be correct. This mapping also is crucial for RNA-like molecules.
References to more rigorous justifications or proofs can be found
below.

\subsection{RNA-like polymers}

\begin{figure}
\centering{}\includegraphics[scale=2]{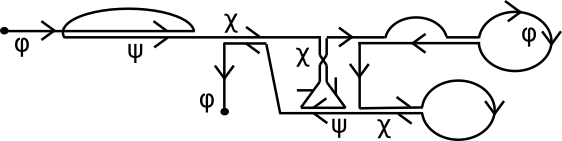}\caption{\label{fig:DoublePolymerDirected}A directed linear polymer $\varphi$
forms a branched polymer with three types of linear polymers, the
single strand polymer $\varphi$ ($\rightarrow$) and double strand
polymers $\chi$ ($\rightleftharpoons$) and $\psi$ ($\Rightarrow$). }
\end{figure}
 One might wonder whether it makes sense to consider even more complex
systems. However, there is one type of system of utmost importance
in biology: RNA molecules (or DNA molecules). Furthermore, RNA molecules
can be created with arbitrary lengths and arbitray base sequences.
One can imagine well-defined physical situations, which are certainly
interesting, even if not directly relevant for biology. To our knowledge
the critical properties of such systems have not been examined from
first principles.

Fig.(\ref{fig:DoublePolymerDirected}) depicts a single strand RNA
molecule $\varphi$, which partially has formed double strands $\chi$
and double strands $\psi$. Every single strand RNA molecule has a
direction (on the ribose backbone), and actually only oppositely aligned
RNA strands bind to a double strand, denoted as $\chi$ in fig.(\ref{fig:DoublePolymerDirected}).
The double strand polymer $\psi$ is unphysical in this sense, but
it is more general to allow both types of double polymer and simply
ascribe different binding energies to $\psi$ and $\chi$.

Deriving field theories from an RNA-like lattice model is nothing
special, also because the terms of the field theory have direct meaning.
At first it appears that directed polymers are mandatory, but as will
become clear below, the most interesting aspects of the problem can
be examined with undirected polymers, here also denoted as $\varphi$
and $\chi$. The point is, that diagrams with the oppositely aligned
$\chi$ are more singular than diagrams with $\psi$, independently
of whether polymer direction is accounted for or ignored in the field
theory. Polymer direction and length variables are required if one
wants to examine RNA molecules with a random base sequence.

The euclidian action of the simplest model variant is $O\left(n\right)$-symmetric
and reads \cite{Dengler2022} 

\begin{align}
S\left(r_{0},\tau_{0},H,\alpha\right) & =\int\mathrm{d}^{d}x\left\{ \tfrac{1}{2}\sum_{\mu=1}^{n}\varphi_{\mu}\left(r_{0}-\nabla^{2}\right)\varphi_{\mu}+\tfrac{1}{2}\sum_{\mu\nu}\chi_{\mu\nu}\left(\tau_{0}-\nabla^{2}\right)\chi_{\mu\nu}\right\} \label{eq:Action_On}\\
 & +\int\mathrm{d}^{d}x\left\{ -g\sum_{\mu\nu}\chi_{\mu\nu}\varphi_{\mu}\varphi_{\nu}-\lambda\sum_{\mu\nu\rho}\chi_{\mu\nu}\chi_{\nu\rho}\chi_{\rho\mu}+\tfrac{\alpha}{2}\bar{\chi}^{2}-H\bar{\chi}\right\} \nonumber \\
 & +\int\mathrm{d}^{d}x\left\{ -g'\bar{\chi}\sum_{\mu}\varphi_{\mu}\varphi_{\mu}-\lambda'\bar{\chi}\sum_{\mu\nu}\chi_{\mu\nu}^{2}-\lambda''\bar{\chi}^{3}\right\} \overset{\mathrm{decouples}}{\longleftarrow}\nonumber \\
 & +\int\mathrm{d}^{d}x\left\{ +u_{1}\sum_{\mu\nu}\varphi_{\mu}^{2}\varphi_{\nu}^{2}+u_{2}\sum\chi_{\mu\nu}^{2}\chi_{\rho\tau}^{2}+u_{3}\sum\varphi_{\mu}^{2}\chi_{\rho\tau}^{2}-w_{2}'\sum\chi_{\mu\nu}^{2}\chi_{\rho\tau}^{2}\chi_{\alpha\beta}^{2}+...\right\} \stackrel[\mathrm{levant}]{\mathrm{irre-}}{\longleftarrow}.\nonumber
\end{align}
Of interest is the $n\rightarrow0$ limit. The last two lines are
only written down for completeness and can be ignored. The symbol
$\bar{\chi}=\sum_{\mu=1}^{n}\chi_{\mu\mu}$ is an abbreviation. The
coupling constant $g$ binds two single strands $\varphi$ to a double
strand $\chi$, the terms with coupling constants $\lambda$ and $\alpha$
are generated from $g$ or $u_{2}$.

Without the (single strand) $\varphi$-field action (\ref{eq:Action_On})
is a variant of the model of Lubensky and Isaacson or Parisi and Sourlas,
and decribes the conventional one-component ($\chi$) branched polymer.
A renormalization group calculation shows, that the $\chi$-sector
of the model retains the properties of the one-component branched
polymer, even in the presence of $\varphi$. Moreover, the critical
exponent $\eta_{\varphi}$ of the $\varphi$ field coincides with
the critical exponent $\eta_{\chi}$ of the $\chi$ field. 

The mapping of the $\chi$-sector of action (\ref{eq:Action_On})
to the Lee-Yang field theory (eq.(\ref{eq:Action_LeeYang}) below)
in two less dimensions is the well-known mapping of conventional branched
polymers \cite{Parisi1981}. The tensor nature of $\chi$ does not
change anything - it suffices to restrict oneself to external fields
in $1,1$-direction and the field component $\chi_{1,1}$.

The origin of the mapping is the $\chi$-propagator in wave vector
representation
\begin{equation}
\left\langle \chi_{\boldsymbol{k}}^{\mu\nu}\chi_{\boldsymbol{p}}^{\rho\tau}\right\rangle =\left(\frac{\delta_{\mu\rho}\delta_{\nu\tau}+\delta_{\mu\tau}\delta_{\nu\rho}}{2v_{k}}-\frac{\alpha\delta_{\mu\nu}\delta_{\rho\tau}}{v_{k}\left(v_{k}+n\alpha\right)}\right)\left(2\pi\right)^{d}\delta^{d}\left(\boldsymbol{k}+\boldsymbol{p}\right),\label{eq:Prop_xi_On}
\end{equation}
where $v_{k}=\tau_{0}+k^{2}$ \footnote{There was a $\tfrac{1}{2}n\alpha$ in ref.(\cite{Dengler2022}) instead
of $n\alpha$. This makes no difference in the limit $n\rightarrow0$.}. It is expedient to think of two propagators, a normal propagator
$1/v_{k}$ and a more singular $-\alpha v_{k}^{-2}$-propagator. The
latter one leads to the dimensional shift, the minus sign correponds
to the imaginary coupling constant in the Lee-Yang field theory \cite{AIM76,Parisi1981}. 

\section{Lee-Yang field theory }

Understanding the Lee-Yang field theory is crucial due to its equivalence
to the chi-sector of action (\ref{eq:Action_On}) in two additional
space dimensions. This field theory is frequently explored with renormalization
group techniques, yet the essential insights can be derived from elementary
considerations. 

For branched polymers one is interested in the case $H=\mathrm{const},$
and the complication of a Fisher renormalization of the critical exponents
and other difficulties occur already in the less complicated Lee-Yang
model. Many calculations are simpler in the Lee-Yang model, and there
are several exact results. 

\begin{figure}
\centering{}\includegraphics[scale=0.5]{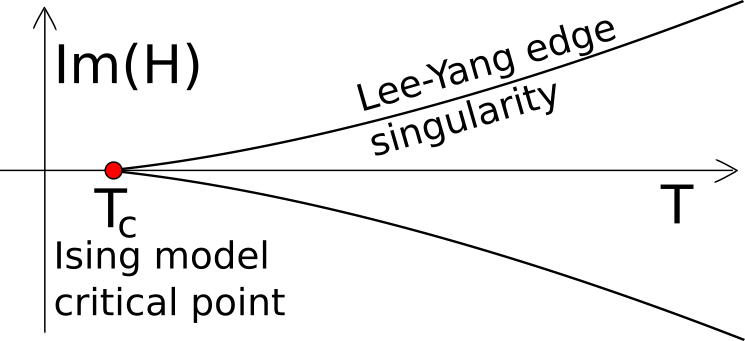}\caption{\label{fig:LeeYangManifold}The critical point of the Ising model
with critical temperature $T_{c}$ ramifies into the two branches
of the Lee-Yang critical manifold in an imaginary magnetic field $H$
for $T>T_{c}.$}
\end{figure}
The starting point is the partition sum of the Ising model, which
has zeros for purely imaginary magnetic field $H$, $\Im\left(H\right)=H''$
and $T>T_{c}$. A region $\left[-H''\left(T\right),H''\left(T\right)\right]$
is without zeros. The curves $\pm H''\left(T\right)$ are a critical
manifold \cite{Fisher1978}, see fig.(\ref{fig:LeeYangManifold}).
The ramification at $T_{c}$ can be described with the Ising model.
The branches for $T>T_{c}$ represent another universality class,
described by the field theory
\begin{equation}
S_{\mathrm{LY}}\left(\tau_{0},H\right)=\int\mathrm{d}^{d}x\left\{ \tfrac{1}{2}\chi\left(\tau_{0}-\nabla^{2}\right)\chi+\lambda\chi^{3}-H\chi\right\} \label{eq:Action_LeeYang}
\end{equation}
with imaginary coupling constant $\lambda$ and imaginary magnetic
field $H$ \cite{Fisher1978}. The field $\chi$ here is a scalar,
there is no $n\rightarrow0$ limit any more, and there is only one
type of propagator. The upper critical dimension of action (\ref{eq:Action_LeeYang})
is six. The critical exponents are known in five-loop order \cite{Borinsky2021},
the values in one and two dimensions are known exactly.

\subsection{Only one parameter}

Replacing the integration variable $\chi$ with $\chi+c$ with constant
imaginary $c$ in the path integral of action (\ref{eq:Action_LeeYang})
leads to
\[
S_{\mathrm{LY}}\left(\tau_{0},H\right)=S_{\mathrm{LY}}\left(\tau_{0}+6\lambda c,H-\tau_{0}c-3\lambda c^{2}\right)+\mathrm{const}.
\]
An appropriate choice of $c$ transforms $\tau_{0}$ to an arbitrary
constant value $\tau_{c},$

\begin{align}
S_{\mathrm{LY}}\left(\tau_{0},H\right) & =S_{\mathrm{LY}}\left(\tau_{c},H+\tfrac{1}{12\lambda}\left(\tau_{0}^{2}-\tau_{c}^{2}\right)\right)+\mathrm{const}.\label{eq:Action_LY_1_Param}
\end{align}
The path integral therefore depends on only one parameter, the second
argument of $S_{\mathrm{LY}}$ on the r.h.s. of eq.(\ref{eq:Action_LY_1_Param}),
and for every $\tau_{0}$ the critical point is reached for some $H=H_{c}$.

This first of all implies that singularities as function of $H$ are
independent of $\tau_{0}$. According to the usual definition of the
critical exponent $\delta$ for instance $\left\langle \chi\right\rangle \sim\left(H-H_{c}\right)^{1/\delta}$. 

Second it is clear, that the critical value of the second argument
also gets traversed linearly if $H$ is held constant and $\tau_{0}$
is varied (except for $\tau_{0}=0$). For constant $H$ thus $\left\langle \chi\right\rangle \sim\tau^{\beta_{H}}$
with $\tau=\tau_{0}-\tau_{0c}$ and $\beta_{H}=1/\delta$. A more
precise expression also includes analytic contributions and reads
$\left\langle \chi\right\rangle \cong a_{0}+a_{1}\tau-A\tau^{\beta_{H}},$
i.e. $\left\langle \chi\right\rangle $ has a peak for $\tau\rightarrow0$
if $0<\beta_{H}\leq1$. The exponent $\beta_{H}$ is in this range
for dimension $d\geq3$, but is negative for $d\leq2$. 

\subsection{One independent critical exponent}

It can be shown that only one of the usual critical exponents $\nu$
and $\eta$ is independent \cite{LubMcKane1981}. Instead of $\eta$
we use $\eta_{\chi}=\eta/2$, the anomalous part of the scaling dimension
of $\chi$.

Consider the classical equation of motion 

\begin{align*}
0 & =\tfrac{1}{Z}\int\mathrm{D}\chi\frac{\delta S_{\mathrm{LY}}}{\delta\chi}e^{-S_{\mathrm{\mathrm{S_{\mathrm{LY}}}}}}=\left\langle \frac{\delta S_{\mathrm{LY}}}{\delta\chi}\right\rangle =\tau_{0}\left\langle \chi\right\rangle +3\lambda\left\langle \chi^{2}\right\rangle -H.
\end{align*}
It is not useful to directly assign a scaling dimension to $\left\langle \chi^{2}\right\rangle $
because of $\left\langle \chi\right\rangle \neq0$. But with $\chi=Q+\chi'$
with $\left\langle \chi'\right\rangle =0$ it follows
\begin{equation}
0=\tau_{0}Q+3\lambda Q^{2}+3\lambda\left\langle \chi'^{2}\right\rangle -H.\label{eq:EqMotion_LY}
\end{equation}
Now assume constant $\tau_{0}.$ The contributions $Q\sim\left(H-H_{c}\right)^{\beta_{H}}$
and $Q^{2}$ are more singular than $H$, but $\left\langle \chi'^{2}\right\rangle $
has positive dimension and consists of analytic and singular contributions.
The analytic contributions compensate $Q$ and $Q^{2}$, and it remains
$\left\langle \chi'^{2}\right\rangle \sim H-H_{c}.$ With the usual
definitions of critical exponents, written as scaling equivalences
for a wavevector $k$,
\begin{equation}
k\sim\left(\chi^{2}\right)^{1/\left(d-1/\nu_{\chi}\right)}\sim\left(H-H_{c}\right)^{1/\left(d/2+1-\eta_{\chi}\right)},\label{eq:ScalingEquivalences}
\end{equation}
it follows
\begin{equation}
\nu_{\chi}=\frac{1}{d/2-1+\eta_{\chi}}.\label{eq:NuFromEta}
\end{equation}
There thus is only one independent critical exponent (one primary
field in the Lee-Yang conformal field theory). For the exponent $\beta_{H}$,
for instance, one finds

\begin{equation}
\beta_{H}=\frac{1}{\delta}=\frac{1}{\left(d/2+1-\eta_{\chi}\right)\nu_{\chi}}=\frac{d-2+2\eta_{\chi}}{d+2-2\eta_{\chi}}.\label{eq:Exp_Beta_H}
\end{equation}

\subsection{Two types of critical exponents}

Action (\ref{eq:Action_LeeYang}) is asymmetric and in general one
has $Q=\left\langle \chi\right\rangle \neq0$. For a perturbation
theory one must set $\chi=Q+\chi'$ with $\left\langle \chi'\right\rangle =0$,
so that the vertex function $\Gamma_{\chi'}$ with a truncated external
$\chi'$ vanishes (no tadpole diagrams). The shift leads to $S_{\mathrm{LY}}\left(\tau_{0}+6\lambda Q,H-\tau_{0}Q-3\lambda Q^{2}\right),$
and in tree approximation $\Gamma_{\chi'}=-H+\tau_{0}Q+3\lambda Q^{2}.$

There now are two possibilities. In the Lee-Yang case one is interested
in $\tau_{0}=\mathrm{const}$,  and $\Gamma_{\chi'}$ drops out with
$Q\sim\left(H-H_{c}\right)^{1/\delta}$ (plus possibly analytic contributions
to $Q$). The fact that $Q$ now plays the role of a temperature variable
in $\tau_{0}+6\lambda Q$ formally corresponds to a critical exponent
$\beta_{\chi}=1$, which is tantamount to relation (\ref{eq:NuFromEta})
between the critical exponents because of the generic equation $\beta_{\chi}/\nu_{\chi}=d/2-1+\eta_{\chi}$.

In the branched polymer case one is interested in $H=\mathrm{const}$,
and $\Gamma_{\chi'}$ drops out with $Q\left(\tau\right)\sim\tau^{\beta_{H}}$
with exponent $\beta_{H}$ from eq.(\ref{eq:Exp_Beta_H}), plus possibly
analytic contributions to $Q\left(\tau\right)$. For $\beta_{H}<1$
then $6\lambda Q$ dominates in the effective temperature variable
$\tau_{0}+6\lambda Q$, and the temperature-dependent critical exponents
get Fisher-renormalized \cite{Fisher1968}. The correlation length
$\xi$, for instance, is $\xi\sim\tau^{-\nu_{H}}$ with $\nu_{H}=\nu\beta_{H}$,
instead of $\xi\sim\tau^{-\nu}$. The suffix $H$ denotes the $H=\mathrm{const}$
critical exponents, relevant for branched polymers. The temperature
independent exponent $\eta_{\chi}$ is not renormalized.

This seems to be somewhat paradoxical in one and two dimensions, where
$\beta_{H}$ is negative, and the effective temperature variable $\tau_{0}+6\lambda Q$\emph{
}diverges\emph{ }for $\tau_{0}\rightarrow\tau_{0c}.$ The difficulty
apparently disappears in a consistent loop expansion, when the loop
expansion of $Q$ also is used instead of the exact value. At any
rate, the exponents $\nu$ and $\gamma$ also become negative in one
and two dimensions, and the effective exponents $\nu_{H}=\nu\beta_{H}$
and $\gamma_{H}=\gamma\beta_{H}$ remain positive.

\begin{onehalfspace}
Here is a list of exactly known Lee-Yang critical exponents 
\begin{equation}
\begin{array}{c|rrrr}
 & \beta_{H} & \eta_{\chi} & \nu_{\chi} & \gamma_{\chi}\\
\hline d=0 & -1\\
d=1 & -\tfrac{1}{2} & -\tfrac{1}{2} & -1 & -3\\
d=2 & -\tfrac{1}{6} & -\tfrac{2}{5} & -\tfrac{5}{2} & -7\\
d=6 & \tfrac{1}{2} & 0 & \tfrac{1}{2} & 1
\end{array}.\label{eq:LY_expExact}
\end{equation}
Of particular interest is the case $d=1$, which corresponds to the
$\chi$-sector of polymer action (\ref{eq:Action_On}) in three space
dimensions.
\end{onehalfspace}

\section{The $O\left(n\right)$-symmetric RNA model in more detail}

\begin{figure}
\centering{}\includegraphics[scale=0.25]{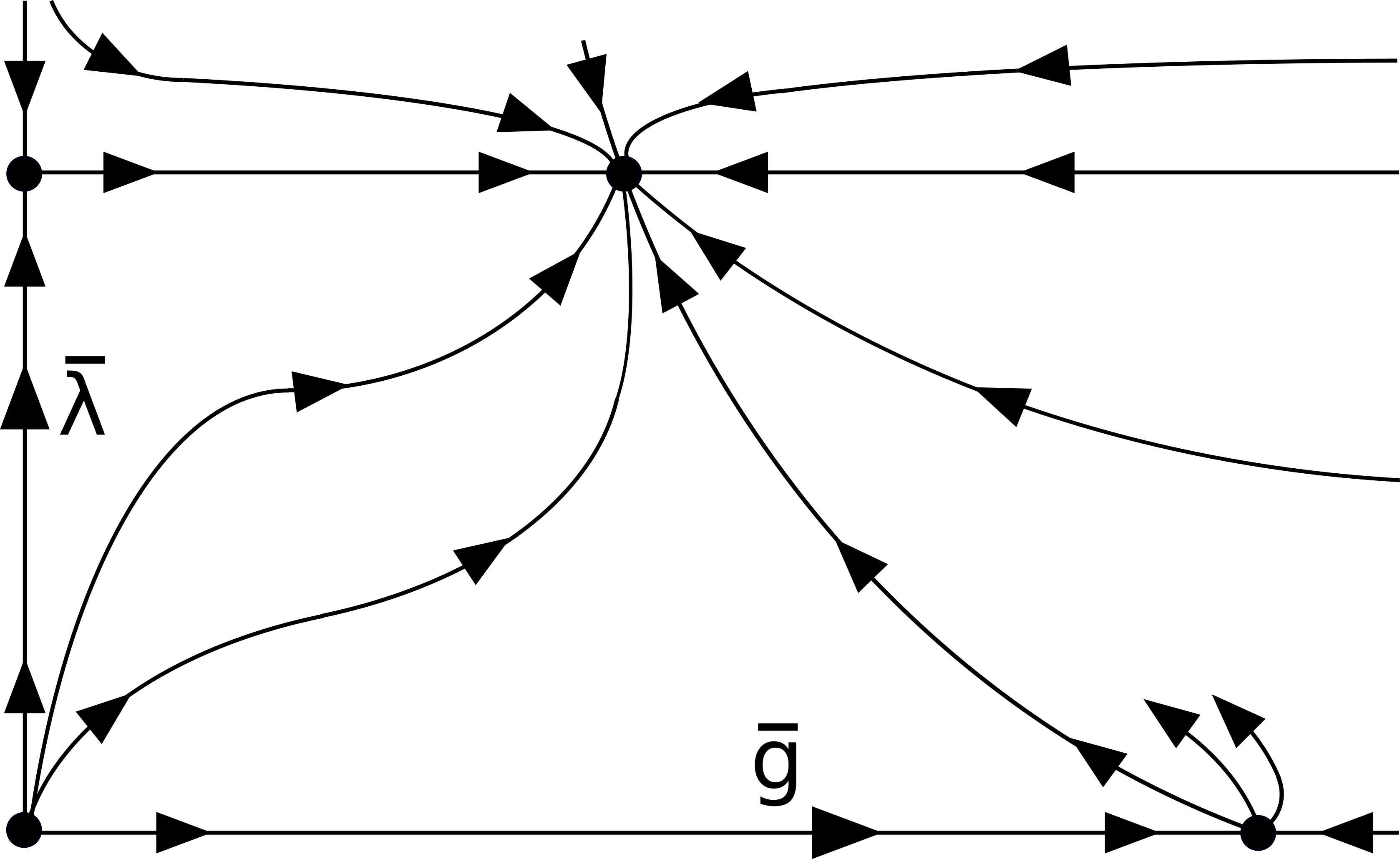}\caption{\label{fig:RG_Flow}The coupling constant flow of model (\ref{eq:Action_On})
for $\alpha>0$, $\epsilon>0$ (schematic).}
\end{figure}
We now examine the $O\left(n\right)$ symmetric RNA model (\ref{eq:Action_On})
in more detail \cite{Dengler2022}. Shifting $\chi$ according to
\begin{equation}
\chi_{\mu\nu}\rightarrow\delta_{\mu\nu}Q+\chi_{\mu\nu}\label{eq:Q_Def_On}
\end{equation}
 in order to eliminate the linear term in $\chi$ gives 
\begin{equation}
S\left(r_{0},\tau_{0},H,\alpha\right)\rightarrow S\left(r_{0}-2gQ,\tau_{0}-6\lambda Q,H-\tau_{0}Q+3\lambda Q^{2},\alpha+4u_{2}Q^{2}\right)+4u_{2}Q\int\mathrm{d}^{d}x\bar{\chi}\sum\chi_{\mu\nu}^{2}+...\label{eq:Action_On_Shift}
\end{equation}
The $\varphi$-sector is only affected by its temperature variable,
$r_{0}\rightarrow r_{\text{0}}-2gQ$, everything else is as for a
one-component ($\chi$) branched polymer \cite{Parisi1981}. In particular,
a $\int\mathrm{d}^{d}x\tfrac{\alpha}{2}\bar{\chi}^{2}$ is generated
if not already present.

Only diagrams with a maximal number of $\chi$-propagators $-\alpha v_{k}^{-2}$
from eq.(\ref{eq:Prop_xi_On}) are of interest. This excludes any
$\varphi$-propagators from the $\chi$-sector. The reason is simple.
Any internal $\varphi$ in a $\Gamma_{\chi\chi}$ or $\Gamma_{\chi\chi\chi}$
vertex function generates a $\varphi$-loop, without a propagator
$-\alpha v_{k}^{-2}$. But the most singular diagrams contain exactly
one propagator $-\alpha v_{k}^{-2}$ per loop \cite{AIM76}, see also
fig.(\ref{fig:DiagramsOn_2Loop}) in the appendix. In other words,
the critical point of the $\chi$-sector is not affected by the single
strand polymer $\varphi$, behaves like a branched polymer of conventional
type, and can be mapped to the Lee-Yang system. This reflects itself
in the one-loop renormalization group flow equations \cite{Dengler2022}

\begin{align}
\mathrm{d}\bar{\lambda}/\mathrm{d}l & =\bar{\lambda}\left(\tfrac{\epsilon}{2}-81\bar{\lambda}^{2}\right),\label{eq:FlowEquFinal}\\
\mathrm{d}\bar{g}/\mathrm{d}l & =\bar{g}\left(\tfrac{\epsilon}{2}-2\bar{g}^{2}-24\bar{g}\bar{\lambda}+9\bar{\lambda}^{2}\right)\nonumber 
\end{align}
for the coupling constants $\bar{\lambda}=\lambda\sqrt{\alpha K_{d}}$
and $\bar{g}=g\sqrt{\alpha K_{d}}$, with $K_{d}=2^{1-d}\pi^{-d/2}/\Gamma\left(d/2\right)$
and $\epsilon=8-d$. The flow of the coupling constant $\bar{\lambda}$
does not depend on the $\chi$-$\varphi$ coupling with coupling constant
$\bar{g}$, see also fig.(\ref{fig:RG_Flow}).

The critical exponents $\eta_{\chi}=-\tfrac{\epsilon}{18}+O\left(\epsilon^{2}\right)$
and $\nu_{\chi}=\tfrac{1}{2}+\tfrac{5}{36}\epsilon+O\left(\epsilon\right)^{2}$
at the stable fixed point $\bar{\lambda}\cong\tfrac{1}{18}\sqrt{2\epsilon}$,
$\bar{g}=\tfrac{1}{6}\sqrt{2\epsilon}$ agree with the Lee-Yang exponents
in $6-\epsilon$ dimensions \cite{Lubensky1979}. For the $\varphi$-field
one finds $\eta_{\varphi}=-\tfrac{\epsilon}{18}+O\left(\epsilon^{2}\right)$.
In fact, in the massless theory there is a hidden symmetry in the
perturbation theory and in the flow equations (\ref{eq:FlowEquFinal})
for $\bar{g}=3\bar{\lambda}$, and it follows $\eta_{\varphi}=\eta_{\chi}$.
This hidden symmetry is illustrated in fig.(\ref{fig:DiagramsOn_2Loop}).
A proof is sketched in the appendix. The symmetry essentially says
that there is no difference between $\varphi$ and $\chi$ at the
stable fixed point.

The exponent $\nu_{\varphi}=\tfrac{1}{2}+\tfrac{\epsilon}{36}+O\left(\epsilon^{2}\right)$,
in contrast, is new. One might think that $\nu_{\varphi}$ could be
related to other exponents with the help of the equation of motion
(see eq.(\ref{eq:NuFromEta})), but this seems not to be the case.

The renormalization group calculations can be done in the massless
theory, but the effective critical exponent $\nu_{H}=$$\beta_{H}\nu_{\chi}$
is the Fisher-renormalized variant.

We now return to the temperature variables $r_{0}$ and $\tau_{0}$
in eq.(\ref{eq:Action_On_Shift}). For $\bar{g}=3\bar{\lambda}$ one
finds $r_{0}\rightarrow r_{0}+6\lambda A\tau^{\beta_{H}}+\mathrm{const}$
and $\tau_{0}\rightarrow\tau_{0}+6\lambda A\tau^{\beta_{H}}+\mathrm{const}$.
There thus is a crossover at $r_{0}-r_{0c}\cong6\lambda A\tau^{\beta_{H}}$.
For small $r_{0}-r_{0c}$ the dependence on $r_{0}$ drops out completely,
and the two field types have the same mass $6\lambda A\tau^{\beta_{H}}.$

\subsection{Mapping of the $\chi$-sector to the Lee-Yang system}

There are some subtle points in the mapping of the $\chi$-sector
of action (\ref{eq:Action_On}) to action (\ref{eq:Action_LeeYang})
in dimension $d-2$ as in \cite{AIM76,Parisi1981}, in particular
for space dimensions $d\leq4$. In fact, the similar mapping of the
random magnetic field Ising model to the normal Ising model in two
less dimensions turned out to be wrong for $d\leq4$ \cite{Rychkov2023}. 

One symptom of a difficulty in the polymer case is related to the
exponent $\beta_{H}<0$ for $d\leq4$. A negative $\beta_{H}$ implies
a diverging $Q$ and diverging effective $r_{0}$-, $\tau_{0}$- and
$\alpha$-values (eq.(\ref{eq:Action_On_Shift})). This difficulty
disappears if one sticks to a consistent loop expansion, when the
expansion of $\beta_{H}$ is used instead of the exact value.

At any rate, in the polymer case the mapping appears to be correct.
As already noted by Parisi and Sourlas there is numerical evidence.
Furthermore, independently of field theory, Brydges and Imbrie \cite{Brydges2003}
proved the equivalence of a particular type of branched polymers with
a hard core gas, which in turn can be mapped to the Lee-Yang system.
More recent references concerning the mapping are \cite{Cardy2003,Rychkov2023}.
We shall not discuss the question any further. It concerns conventional
branched polymers, and is not specific to the RNA model (\ref{eq:Action_On}).

\section{Field theories with length variables}

Directed polymers with length variables are useful to understand how
the $O\left(n\right)$ model, which has directionless polymers, can
describe RNA molecules with an internal direction (only oppositely
aligned RNA strands bind to a double strand). Furthermore, the model
with length variables allows to examine the effect of a random base
sequence.

A derivation of polymer field theories with length variables $s$
instead of the $O\left(n\right)$ indices (as in eq.(\ref{eq:Action_On}))
can be found in ref.(\cite{Dengler2020}) and in section (\ref{subsec:FieldTheoryDeriv})
below. When ``frequencies'' are used instead of length variables,
then these frequencies flow unchanged from external sources to external
sinks, exactly like the $O\left(n\right)$-indices. There are no frequency
integrals, and the calculations are similar to the $O\left(n\right)$-model,
but without the $n\rightarrow0$ limit. Closed loops drop out because
length variables cannot increase consistently in a loop. 

The direction of increasing length variable $s$ gives the single
strand $\varphi$ ($\rightarrow$) and the double strand polymers
$\psi$ ($\Rightarrow$) a direction, and there actually are pairs
of fields, $\varphi\left(\boldsymbol{x},s\right)$, $\tilde{\varphi}\left(\boldsymbol{x},s\right)$
and $\psi\left(\boldsymbol{x},s,s'\right)$, $\tilde{\psi}\left(\boldsymbol{x},s,s'\right)$.
If double polymer twisting is ignored (which is justified for long
polymers), then $\psi$ and $\tilde{\psi}$ are symmetric in their
length variables (polymer twisting is difficult to define for $d\geq3$
anyway). The double strand polymer $\chi$ ($\rightleftharpoons$)
with oppositely aligned single strands has no internal direction,
and requires only one field $\chi$, but with $\chi\left(\boldsymbol{x},s,s'\right)\neq\chi\left(\boldsymbol{x},s',s\right)$. 

The (euclidian) action for the RNA-like problem in frequency representation
reads (variables conjugate to a length variable under Fourier transformation
are called ``frequencies'' $\omega$, $\omega'$, $\omega''$)

\begin{align}
S & =S_{\varphi}+S_{\chi}+S_{\varphi\chi},\label{eq:Action_Lengths}\\
S_{\varphi} & =\int\mathrm{d}^{d}x\left\{ \int_{\omega}\tilde{\varphi}_{-\omega}\left(r_{0}-\nabla^{2}-i\omega\right)\varphi_{\omega}+u_{\varphi}\left(\int_{\omega}\tilde{\varphi}_{-\omega}\varphi_{\omega}\right)^{2}\right\} ,\nonumber \\
S_{\chi} & =\int\mathrm{d}^{d}x\left\{ \tfrac{1}{2}\int_{\omega,\omega'}\chi_{-\omega',-\omega}\left(\tau_{0}-\nabla^{2}-i\left(\omega-\omega'\right)\right)\chi_{\omega,\omega'}+\tfrac{\alpha}{2}\left(\int_{\omega}\chi{}_{\omega,-\omega}\right)^{2}-H\int_{\omega}\chi_{\omega,-\omega}\right\} \nonumber \\
 & \qquad+\int\mathrm{d}^{d}x\left\{ -\lambda\int_{\omega,\omega',\omega''}\chi_{\omega,-\omega'}\chi_{\omega',-\omega''}\chi_{\omega'',-\omega}+u_{\chi}\left(\int_{\omega,\omega'}\chi_{-\omega',-\omega}\chi_{\omega,\omega'}\right)^{2}\right\} ,\nonumber \\
S_{\varphi\chi} & =\int\mathrm{d}^{d}x\left\{ -2g\int_{\omega,\omega'}\tilde{\varphi}_{\omega}\varphi_{\omega'}\chi_{-\omega,-\omega'}+u_{\varphi\chi}\left(\int_{\omega}\tilde{\varphi}_{-\omega}\varphi_{\omega}\right)\left(\int_{\omega,\omega'}\chi_{-\omega',-\omega}\chi_{\omega,\omega'}\right)\right\} .\nonumber 
\end{align}
For simplicity all space arguments of the fields are suppressed.
The ``frequency'' integrals $\int_{\omega}=\int\mathrm{d}\omega/\left(2\pi\right)$
look complicated, but frequencies run unchanged from source to sink
through Feynman diagrams, exactly like the indices of $O\left(n\right)$
models. 

The action $S_{\varphi}$ describes a directed polymer ($\rightarrow$).
The action $S_{\chi}$ with $\lambda=0$ and $\alpha=0$ describes
a linear polymer ($\rightleftharpoons$) with two length variables,
i.e. in effect a polymer without direction. The complete action $S_{\chi}$
describes a conventional branched polymer. The action $S$ in total
is the RNA model with only oppositely aligned double strand polymers,
the excluded volume interactions with coupling constants $u_{\varphi}$,
$u_{\chi}$ and $u_{\varphi\chi}$ are irrelevant at the critical
dimension $8$.

The double polymer $\psi$ ($\Rightarrow$) is of minor interest,
there are no such RNA double strands. But for completeness,
\begin{align*}
S_{\psi} & =\int\mathrm{d}^{d}x\left\{ \int_{\omega,\omega'}\tilde{\psi}_{-\omega,-\omega'}\left(\hat{\tau}_{0}-\nabla^{2}-i\hat{w}\left(\omega+\omega'\right)\right)\psi_{\omega,\omega'}-\hat{g}\int_{\omega,\omega'}\left(\tilde{\psi}_{\omega,\omega'}\varphi_{-\omega}\varphi_{-\omega'}+\tilde{\varphi}_{\omega}\tilde{\varphi}_{\omega'}\psi_{-\omega,-\omega'}\right)\right\} \\
 & \qquad-\hat{\lambda}\int_{\omega,\omega',\omega''}\chi_{\omega,\omega'}\tilde{\psi}_{\omega'',-\omega}\psi_{-\omega'',-\omega'}+u_{\psi}\int\mathrm{d}^{d}x\left(\int_{\omega,\omega'}\tilde{\psi}_{-\omega,-\omega'}\psi_{\omega,\omega'}\right)^{2}+...
\end{align*}
The action $S_{\psi}+S_{\varphi}$ with $\hat{\lambda}=0$ is a model
with only $\Rightarrow$ double polymers without a condensate $Q$,
the critical dimension is $6$ \cite{Dengler2020}.

\subsection{Relation to $O\left(n\right)$-symmetric model}

It is now possible to repeat the calculations done with the $O\left(n\right)$
action integral (\ref{eq:Action_On}) with action integral (\ref{eq:Action_Lengths}).
We only sketch some points. The propagators are

\begin{align}
\left\langle \tilde{\varphi}_{\omega'}^{\boldsymbol{p}}\varphi_{\omega}^{\boldsymbol{k}}\right\rangle  & =\frac{\delta\left(\omega+\omega'\right)}{r_{0}+k^{2}-i\omega}\left(2\pi\right)^{d+1}\delta^{d}\left(\boldsymbol{p}+\boldsymbol{k}\right),\label{eq:Prop_Chi_omega}\\
\left\langle \chi_{\omega,\omega'}^{\boldsymbol{p}}\chi_{\nu,\nu'}^{\boldsymbol{k}}\right\rangle  & =\left(\frac{\delta\left(\omega+\nu'\right)\delta\left(\omega'+\nu\right)}{v_{k}-i\left(\omega-\omega'\right)}-\alpha\frac{\delta\left(\omega+\omega'\right)}{v_{k}-i\left(\omega-\omega'\right)}\frac{\delta\left(\nu+\nu'\right)}{v_{k}-i\left(\nu-\nu'\right)}\right)\left(2\pi\right)^{d+2}\delta^{d}\left(\boldsymbol{p}+\boldsymbol{k}\right),\nonumber 
\end{align}
with $v_{k}=\tau_{0}+k^{2}$, and the analogy to the $O\left(n\right)$-model
propagator (\ref{eq:Prop_xi_On}) is clear.

To get rid of terms linear in $\chi$ one shifts $\chi$ according
to
\begin{equation}
\chi_{\omega,\omega'}\rightarrow2\pi\delta\left(\omega+\omega'\right)Q\left(\tau_{0},\omega\right)+\chi{}_{\omega,\omega'}.\label{eq:Q_Def_Lengths}
\end{equation}
This equation is the counterpart to eq.(\ref{eq:Q_Def_On}). The condensate
$Q$ now depends on $\tau_{0}$ and $\omega$. The shift reproduces
eq.(\ref{eq:Action_On_Shift}), except that the linear term
\[
\int\mathrm{d}^{d}x\int_{\omega}\left(-H-3\lambda Q^{2}+\left(\tau_{0}-2i\omega\right)Q\right)\chi_{\omega,-\omega}
\]
now contains $\tau_{0}-2i\omega$ instead of $\tau_{0}$. 

To compare the renormalization group calculation it suffices to determine
the combinatoric factors of the diagrams. We have done this in one-loop
order and find the same factors. The coupling constant $2g$ in $S_{\varphi\chi}$
compensates a factor $2$, which comes from the square in $g\chi\varphi^{2}$
in the $O\left(n\right)$ action (\ref{eq:Action_On}). 

It is also clear that the double strand polymer $\psi\left(\Rightarrow\right)$
does not contribute to the $\chi$ sector. Any internal $\psi/\tilde{\psi}$
in a $\Gamma_{\chi\chi}$ or $\Gamma_{\chi\chi\chi}$ vertex function
either generates a $\psi$-loop, without a propagator $-\alpha v_{k}^{-2}$,
or terminates at a $\psi\tilde{\varphi}^{2}$ or a $\tilde{\psi}\varphi^{2}$,
which generates a $\varphi$-loop, also without a propagator $-\alpha v_{k}^{-2}$.
But the most singular diagrams of the $\chi$ sector require exactly
one such propagator per loop \cite{AIM76}, see fig.(\ref{fig:DiagramsOn_2Loop})
in the appendix.

This confirms that the $O\left(n\right)$ symmetric action (\ref{eq:Action_On})
with undirected polymers describes a RNA-like situation, where only
oppositely aligned single strands form double strands.

\subsection{Derivation of the $\chi$ field theory from a lattice model}

\label{subsec:FieldTheoryDeriv}%
Here we derive the $\chi$-sector ($\rightleftharpoons$) of eq.(\ref{eq:Action_Lengths})
from a lattice model. The derivation in ref.(\cite{Dengler2022})
resulted in a field theory with a pair of fields, one of which could
be eliminated. Instead, we take a different approach and immediately
start with a single type of operator and field.

Assume that for every lattice point $i$ there are operators $C_{i}^{\mu,\nu}$
carrying two length indices, $\mu,\nu\in\mathbb{Z}$. All operators
commute and $C_{i}^{\mu\nu}C_{i}^{\rho\tau}C_{i}^{\beta\gamma}=0$
for any length indices. There is an ``average'' with $\left\langle C_{i}^{\mu\nu}\right\rangle =0$
and
\begin{equation}
\left\langle C_{i}^{\mu\nu}C_{j}^{\rho\tau}\right\rangle =\delta_{ij}\delta_{\mu\tau}\delta_{\nu\rho}.\label{eq:AvgOper_C_C}
\end{equation}
The transposition of the length indices in eq.(\ref{eq:AvgOper_C_C})
is essential. Define a partition sum

\begin{equation}
Z=\left\langle \prod_{i,j;\mu,\nu}e^{\tfrac{1}{2}C_{i}^{\nu+1,\mu-1}v_{i,j}C_{j}^{\mu,\nu}}\right\rangle =\left\langle \prod_{i,j;\mu,\nu}\left(1+\tfrac{1}{2}C_{i}^{\nu+1,\mu-1}v_{i,j}C_{j}^{\mu,\nu}\right)\right\rangle ,\label{eq:Z_chi_Lengths}
\end{equation}
where $v_{i,j}=1$ for next neighbor lattice points and $v_{i,j}=0$
otherwise. Only the constant and linear part of the expansion of the
exponential function contribute because of $0=C_{i}^{3}=\left\langle \left(C_{j}^{\mu,\nu}\right)^{2}\right\rangle $.
To understand the meaning of $Z$ it suffices to consider the factors
containing lattice point $0$. If $m$ and $n$ denote the neighbors
of lattice point $0$ then

\begin{align*}
Z & =\left\langle ...\prod_{m;\mu,\nu}\left(1+C_{0}^{\nu+1,\mu-1}C_{m}^{\mu,\nu}\right)\prod_{n;\rho,\tau}\left(1+C_{n}^{\tau+1,\rho-1}C_{0}^{\rho,\tau}\right)\right\rangle \\
 & =\left\langle ...\prod_{m,n;\mu,\nu,\rho,\tau}\left(1+C_{0}^{\nu+1,\mu-1}C_{0}^{\rho,\tau}C_{m}^{\mu,\nu}C_{n}^{\tau+1,\rho-1}\right)\right\rangle =\left\langle ...\prod_{m,n;\mu,\nu}\left(1+C_{n}^{\nu+2,\mu-2}C_{m}^{\mu,\nu}\right)\right\rangle .
\end{align*}
Here it has been used that $C_{0}$ must occur in pairs in the average.
The remaining expression has the same form as the original one, except
that $C_{0}$ is missing and there are mutually exclusive paths from
neighbor $m$ to lattice point $0$ to neighbor $n$, in which one
length variable increments by two and the other length variable decrements
by two. The case $m=n$ would be a closed loop, with inconsistent
length variables. In total, $Z$ counts all self-avoiding paths of
double strands of the $\rightleftharpoons$-type. All closed loops
contain inconsistent length variables, and to have any real paths
requires external sources and sinks, factors in $Z$ like $1+C_{i}^{\mu\nu}$.

The next step is a Hubbard-Stratonovich transformation. Let us write
$Z=\left\langle e^{\tfrac{1}{2}CVvC}\right\rangle $, where $V^{\mu\nu;\rho\tau}=\delta_{\mu,\tau+1}\delta_{\nu,\rho-1}$
is a  $\mathbb{Z}^{2}\times\mathbb{Z}^{2}$-matrix with index pairs
$\mu\nu$ and $\rho\tau$ and $V=V^{-1}$.  This matrix is symmetric
in the sense $V^{\mu\nu;\rho\tau}=V^{\rho\tau;\mu\nu}$, and it is
to be shown

\begin{equation}
e^{\tfrac{1}{2}CVvC}=\int_{\mathscr{D}}\mathrm{D}\chi e^{-\tfrac{1}{2}\chi\left(Vv\right)^{-1}\chi+\chi\cdot C},\label{eq:HubbStrato}
\end{equation}
where $\chi\cdot C=\sum\chi_{i}^{\mu\nu}C_{i}^{\mu\nu}$. This equation
looks like a standard gaussian integral, but $V$ is not positive
definite (eigenvalues are $\pm1$), and integration along $\mathbb{R}$
is not sufficient. However, it is not difficult to specify an appropriate
integration domain $\mathscr{D}$. For the invariant operators $C^{\mu,\mu-1}$
defined by $VC=C$ integrating $\chi^{\mu,\mu-1}$ along $\mathbb{R}$
in eq.(\ref{eq:HubbStrato}) reproduces the l.h.s..

The other quantities $\left(C,C'\equiv VC\right)$ and $\left(\chi,\chi'\equiv V\chi\right)$
are paired, and the sum in the exponent in eq.(\ref{eq:HubbStrato})
can be restricted to one component,
\[
{\textstyle \sum}{}^{\left(1\right)}\left(-\chi v^{-1}\chi'+\chi C+\chi'C'\right).
\]
Integrating $\chi$ along $i\mathbb{R}$ now generates a function
$\delta$$\left(\chi'-vC\right)$. Integrating $\chi'$ along $\mathbb{R}$
then converts the remaining exponent to ${\textstyle \sum}{}^{\left(1\right)}\chi'C'={\textstyle \sum}{}^{\left(1\right)}\left(vC\right)C'=\tfrac{1}{2}CVvC$,
as in eq.(\ref{eq:HubbStrato}). In total, the field $\chi$ is to
be decomposed into invariant part $\chi^{\mu,\mu-1}$ with integration
domain $\mathbb{R}$ and pairs ($\chi,V\chi$) with integration domain
$\left(i\mathbb{R},\mathbb{R}\right)$.

As usual the remaining average $\left\langle e^{\chi\cdot C}\right\rangle $
in eq.(\ref{eq:HubbStrato}) can be calculated, with the result

\[
Z=\int_{\mathscr{D}}\mathrm{D}\chi e^{-\tfrac{1}{2}\chi v^{-1}V\chi+\sum_{i}\ln\left(1+\tfrac{1}{2}\sum_{\mu\nu}\chi_{i}^{\mu\nu}\chi_{i}^{\nu\mu}\right)}.
\]
This is the $\chi$-sector of eq.(\ref{eq:Action_Lengths}) with
discrete length and space variables and repulsive $\chi^{4}$-interaction.
The interaction $-g\chi\tilde{\varphi}\varphi$ can be added directly
to the field theory, or at operator level in eq.(\ref{eq:Z_chi_Lengths}).
The interaction $-\lambda\chi^{3}$ is generated from $-g\chi\tilde{\varphi}\varphi$
and need not be added explicitly (there is no strictly local operator
expression for $-\lambda\chi^{3}$ because of $C_{i}^{3}=0$, but
a quasi-local expression would suffice). 

\section{RNA with random base sequence}

The base sequence of real RNA is neither random nor deterministic,
but a random base sequence is a reasonable approximation. A random
base sequence can be examined with action (\ref{eq:Action_Lengths})
with length variables, not with the $O\left(n\right)$ symmetric action
(\ref{eq:Action_On}). 

A simple physical quantity is the probability $P\left(\boldsymbol{x}\right)\mathrm{d}^{d}x$
to find the end of a $\varphi$-polymer in the vicinity of point $\boldsymbol{x}$,
if its start point is at the origin. If $\left[...\right]_{*}$ denotes
the base sequence average and $\left\langle ...\right\rangle $ the
path integral, then the answer is contained in the generating functional

\begin{align}
J\left(h\right) & =\left[\ln\left\langle \tilde{\varphi}\left(0,0\right)\int\mathrm{d}^{d}x\mathrm{d}sh\left(\boldsymbol{x}\right)\varphi\left(\boldsymbol{x},s\right)\right\rangle \right]_{*}\label{eq:GenFunc_Random}
\end{align}
according to 
\[
P\left(\boldsymbol{x}\right)=\left.\frac{\delta}{\delta h\left(\boldsymbol{x}\right)}J\left(h\right)\right|_{h=1}=\left[\frac{\left\langle \tilde{\varphi}\left(0,0\right)\int\mathrm{d}s\varphi\left(\boldsymbol{x},s\right)\right\rangle }{\int\mathrm{d}^{d}x\mathrm{d}s\left\langle \tilde{\varphi}\left(0,0\right)\varphi\left(\boldsymbol{x},s\right)\right\rangle }\right]_{*}.
\]
The $\varphi$ propagator is not properly normalized, and it is essential
to normalize it before performing the base sequence average (one also
could omit the $s$-integral to obtain the probability to find the
end of a polymer with given length $s$ at $\boldsymbol{x}$). As
usual one uses the replica trick, and instead of calculating $\ln\left\langle ...\right\rangle $
one calculates 

\[
\left\langle ...\right\rangle ^{n}=\int\prod_{r=1}^{R}\left\{ \mathrm{D}\varphi_{r}\mathrm{D\tilde{\varphi}_{r}}\mathrm{D}\chi_{r}\tilde{\varphi}_{r}\left(0,0\right)\int_{x}\mathrm{d}sh\left(\boldsymbol{x}\right)\varphi_{r}\left(\boldsymbol{x},s\right)\right\} e^{-\sum_{r=1}^{R}S\left(\varphi_{r},\tilde{\varphi}_{r},\chi_{r}\right)},
\]
for arbitrary $r\in\mathbb{N}$, and then uses the analytic continuation
$\ln\left\langle ...\right\rangle =\lim_{R\rightarrow0}\tfrac{1}{R}\left(\left\langle ...\right\rangle ^{R}-1\right).$
Every replica contains a $\varphi$-polymer, which can form double
polymers as in fig.(\ref{fig:DoublePolymerDirected}).

\subsection{Base sequence average}

In the case of a random base sequence the (negative) binding energy
$\tau_{0}$ of the double polymer in eq.(\ref{eq:Action_Lengths})
is to be replaced with a random $\tau_{0}\left(s,s'\right)$, depending
on the two length variables. Returning to discrete length variables
$\mu,\mu'\in\mathbb{N}_{0}$ instead of $s$,$s'$ allows to write
the base sequence as $b_{\mu}\in\left\{ 0,1,2,3\right\} $, with each
$b_{\mu}$ drawn from the probability distribution $p\left(b\right)=\sum_{r=0}^{3}p_{r}\delta_{b,r}$,
with probabilities $\sum_{r=0}^{3}p_{r}=1$. The (negative) binding
energy $\tau_{0}\left(\mu,\mu'\right)\equiv\tau\left(b_{\mu},b_{\mu'}\right)$
then is a symmetric $4\times4$ matrix. The base sequence average
is

\[
\left[e^{-S}\right]_{*}=\left(\prod_{\mu}\sum_{b_{\mu}}p\left(b_{\mu}\right)\right)e^{-\tfrac{1}{2}\int_{x}\sum_{r,\mu,\mu'}\chi_{\mu',\mu}^{r}\tau\left(b_{\mu},b_{\mu'}\right)\chi_{\mu,\mu'}^{r}}\equiv\left[e^{X}\right]_{*}=e^{\kappa_{1}\left(X\right)+\tfrac{1}{2!}\kappa_{2}\left(X\right)+...},
\]
where $\kappa_{r}\left(X\right)$ are the cumulants of $X$. One calculates

\begin{align*}
\kappa_{1} & =\left[X\right]_{*}=-\tfrac{1}{2}\int_{x}\sum_{r,\mu,\mu'}\chi_{\mu',\mu}^{r}\left[\tau\left(b_{\mu},b_{\mu'}\right)\right]_{*}\chi_{\mu,\mu'}^{r}.
\end{align*}
The average $\left[...\right]_{*}$ in $\kappa_{1}$ is some constant,
which can be interpreted as $\tau_{0}$. Now consider

\[
\kappa_{2}=\left[X^{2}\right]_{*}-\left[X\right]_{*}^{2}=\tfrac{1}{4}\left(\prod_{\mu}\sum_{b_{\mu}}p\left(b_{\mu}\right)\right)\left(\int_{x}\sum_{r,\mu,\mu'}\chi_{\mu',\mu}^{r}\tau\left(b_{\mu},b_{\mu'}\right)\chi_{\mu,\mu'}^{r}\right)^{2}-\left[X\right]_{*}^{2}.
\]
The factors $\left(\int_{x}\sum_{r,\mu,\mu'}...\right)\left(\int_{x'}\sum_{r',\nu,\nu'}...\right)$
are correlated for $\mu=\nu$, $\mu'=\nu'$, $\mu=\nu'$, $\mu'=\nu$.
 Averages like $\left[\tau\left(b_{\mu},b_{\mu'}\right)\tau\left(b_{\mu},b_{\nu}\right)\right]_{*}$
with one common length index (common base) give the same constant.
Thus

\[
\kappa_{2}^{\left(1\right)}=\lambda_{2}\int_{x,x'}\sum_{r,r',\mu,\mu',\nu}\chi_{\mu',\mu}^{r}\left(\boldsymbol{x}\right)\chi_{\mu,\mu'}^{r}\left(\boldsymbol{x}\right)\chi_{\nu,\mu}^{r'}\left(\boldsymbol{x}'\right)\chi_{\mu,\nu}^{r'}\left(\boldsymbol{x}'\right).
\]
Length indices can also be pairwise identical. This leads to
\[
\kappa_{2}^{\left(2\right)}=\hat{\lambda}_{2}\int_{x,x'}\sum_{r,r',\mu,\mu'}\chi_{\mu',\mu}^{r}\left(\boldsymbol{x}\right)\chi_{\mu,\mu'}^{r}\left(\boldsymbol{x}\right)\chi_{\mu',\mu}^{r'}\left(\boldsymbol{x}'\right)\chi_{\mu,\mu'}^{r'}\left(\boldsymbol{x}'\right)
\]
with one length sum less.

\subsection{Effective action}

In frequency representation the interactions are  
\begin{align}
S_{2} & =-\lambda_{2}\int_{x,x'}\sum_{r,r'}^{R}\int_{\omega,\omega',\nu,\nu',\rho}\chi_{-\omega',-\omega}^{r}\left(\boldsymbol{x}\right)\chi_{\omega+\rho,\omega'}^{r}\left(\boldsymbol{x}\right)\chi_{-\nu',-\nu}^{r'}\left(\boldsymbol{x}'\right)\chi_{\nu-\rho,\nu'}^{r'}\left(\boldsymbol{x}'\right)\label{eq:Interaction_Random}\\
 & \quad-\hat{\lambda}_{2}\int_{x,x'}\sum_{r,r'}^{R}\int_{\omega,\omega',\nu,\nu',\rho,\rho'}\chi_{-\omega',-\omega}^{r}\left(\boldsymbol{x}\right)\chi_{\omega+\rho,\omega'+\rho'}^{r}\left(\boldsymbol{x}\right)\chi_{-\nu',-\nu}^{r'}\left(\boldsymbol{x}'\right)\chi_{\nu-\rho,\nu'-\rho'}^{r'}\left(\boldsymbol{x}'\right).\nonumber 
\end{align}
To avoid confusion, $\omega$, $\nu$, and $\tau$ denote ``frequencies''
in eq.(\ref{eq:Interaction_Random}). These interactions couple the
replicas, and transfer frequencies but no wavevector between two double
polymer strands $\rightleftharpoons$. The coupling constant $\lambda_{2}$
has wavevector dimension $\left[\lambda_{2}\right]=2$ in any space
dimension and is nominally strongly relevant, the coupling constant
$\hat{\lambda}_{2}$ is dimensionless. %

Higher order cumulants have the same dimension, and in total there
results a rather complicated problem. The conclusion is that the critical
behavior found for periodic base sequences breaks down for random
base sequences.

\section{Conclusions}

It has been shown that there are precisely defined field theories
for RNA-like polymers in a good solvent. These field theories can
be derived from lattice models, and consequently should exactly describe
the critical properties of such systems.

We now summarize the case of a periodic base sequence, that is constant
pairing energy. The $O\left(n\right)$-symmetric action (\ref{eq:Action_On})
and action (\ref{eq:Action_Lengths}) with directed polymers and length
variables give the same result: at the stable fixed point the double-strand
sector ($\chi$) of the models decouples from the single-strand sector,
and becomes a conventional branched polymer (lattice animal), which
by a well-known mapping is equivalent to the Lee-Yang field theory
(\ref{eq:Action_LeeYang}) in two less dimensions. 

In the $O\left(n\right)$ symmetric model there is no distinction
between double polymers of type $\rightleftharpoons$ and of type
$\Rightarrow$. It nevertheless produces the same result. The point
is, that the diagrams are still there, but diagrams with RNA-like
double polymers $\rightleftharpoons$ are more singular.

To our knowledge the predictions for conventional branched polymers
have only been compared with series expansions and Monte Carlo simulations,
not with experiment \cite{Parisi1981}. RNA molecules with given length
and periodic base sequence might be a precisely defined candidate
for an experiment. The gyration radius of branched RNA molecules should
increase with length $\ell$ like $\ell^{1/2}$ \cite{Parisi1981}.

The single strand polymer $\varphi$ does not change the critical
behavior of the double strand polymer. However, it gets influenced
by the double strand polymer and contributes a new critical exponent
$\nu_{\varphi}$. This exponent can be calculated in an expansion
around the critical dimension $d_{c}=8$, but there is a large gap
between $d_{c}=8$ and three dimensions. In the final stage of pair
condensation we find a power law $s_{\varphi}\sim\tau^{Z}$ for the
remaining amount of single strand polymer, with an exponent $-2/\nu_{\varphi}$
in three dimensions. It would be of interest to obtain an accurate
value for the critical exponent $\nu_{\varphi}$ in three and two
dimensions. This would require at least a five-loop calculation, as
for the similar exponent $\nu_{\chi}$ \cite{Borinsky2021}. Series
expansions might be an alternative. At the moment it even is not clear
whether $\nu_{\varphi}$ really is negative for $d\leq4$. 

We have also investigated the impact of a random RNA base sequence.
This is only possible within the more detailed field theory with polymer
length variables. The randomness generates complicated relevant interactions,
rendering the conclusions drawn from the model with a periodic base
sequence invalid.\pagebreak{}

\bibliographystyle{habbrv}
\bibliography{DoublePolymer_Synopsis}

\pagebreak{}

\appendix

\section*{Appendix}

\addcontentsline{toc}{section}{Appendix}

\begin{figure}[h]
\centering{}\includegraphics[scale=1.6]{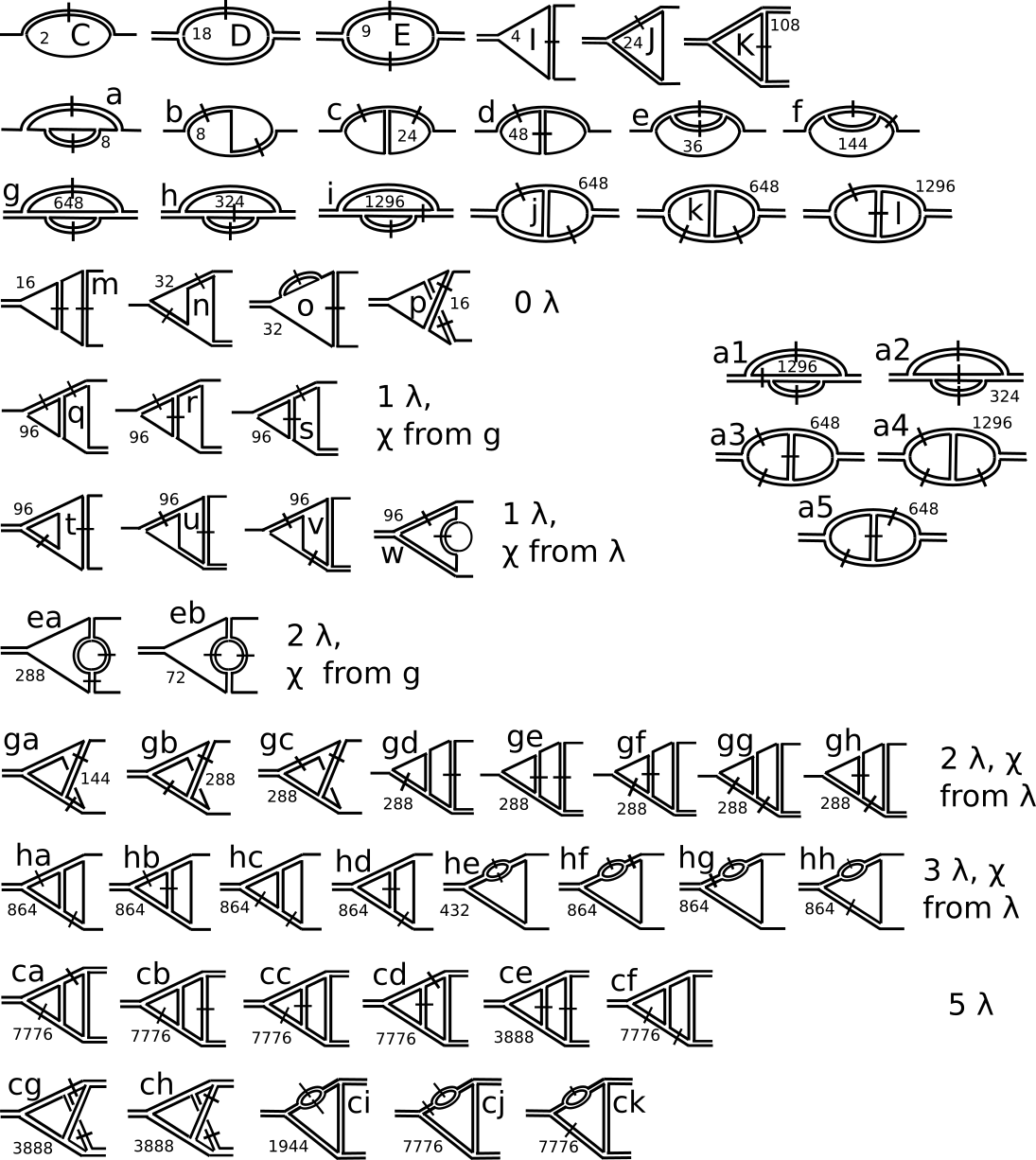}\caption{\label{fig:DiagramsOn_2Loop}One- and two-loop diagrams and their
combinatoric factors for $O\left(n\right)$ symmetric action (\ref{eq:Action_On}).
The bar denotes the propagator proportional to $v_{k}^{-2}$ from
eq.(\ref{eq:Prop_xi_On}).}
\end{figure}
There is a hidden symmetry between $\varphi$- and $\chi$- diagrams
in the massless theory at the stable fixed point with $g=3\lambda.$
The combinatoric factors of diagram $e$ ($36$) and $h$ ($324$),
for instance, produce the same value, because $3^{2}36_{e}=324_{h}.$
Other examples are $3^{3}4_{I}+3^{2}24_{J}=3\cdot108_{K}$ and $3^{5}16_{p}+3^{3}288_{gc}=3\cdot3888_{ch}$.

Here is a way to get the $\Gamma_{\varphi\varphi}$ and $\Gamma_{\chi\varphi\varphi}$
vertex functions from the vertex functions $\Gamma_{\chi\chi}$ and
$\Gamma_{\chi\chi\chi}$. Imagine that a diagram for $\Gamma_{\chi\chi}$
or $\Gamma_{\chi\chi\chi}$ is cut at the $v_{k}^{-2}$ propagators
(the bar). There results a tree diagram, with a unique path between
any two external lines. Select two external lines (three possibilities
in the $\Gamma_{\chi\chi\chi}$ case), and replace them and the path
between them with $\varphi$. This gives a diagram for $\Gamma_{\varphi\varphi}$
or $\Gamma_{\chi\varphi\varphi}$. The combinatoric factors of the
diagrams are related. If originally there is a $\lambda^{n}$ and
there are $k$ vertices on the path then perturbation theory starts
with a factor $\lambda^{n}\tfrac{1}{n!}\binom{n}{k}=\lambda^{n}\tfrac{1}{\left(n-k\right)!}\tfrac{1}{k!}$
when $k$ vertices $\lambda$ are singled out for the path. The contraction
along the path gives a factor $3$ for every $\lambda.$ This is to
be compared with $\lambda^{n-k}\tfrac{1}{\left(n-k\right)!}g^{k}\tfrac{1}{k!}$
for the $\Gamma_{\varphi\varphi}$ or $\Gamma_{\chi\varphi\varphi}$
diagram. 
\end{document}